\documentclass[11pt]{article}
\usepackage{moriond,epsfig}

\def\Journal#1#2#3#4{{#1} {\bf #2}, #3 (#4)}

\def\NIMA{{\em Nucl. Instrum. Methods} A}

\def\PLB{{\em Phys. Lett.}  B}
\def\PRL{\em Phys. Rev. Lett.}

\def\be{\begin{equation}}
\def\ee{\end{equation}}
\def\bea{\begin{eqnarray}}
\def\eea{\end{eqnarray}}

\def\bckpp{B^\pm\to K^\pm\pi^+\pi^-}
\def\bckkk{B^\pm\to K^\pm K^+K^-}
\def\bpkpp{B^+\to K^+\pi^+\pi^-}
\def\bpkkk{B^+\to K^+K^+K^-}
\def\kckk{K^\pm K^+K^-}
\def\kcpp{K^\pm\pi^+\pi^-}
\def\kppp{K^+\pi^+\pi^-}
\def\kmpp{K^-\pi^-\pi^+}
\def\kpkk{K^+K^+K^-}

\def\de{\Delta E}
\def\mb{M_{bc}}
\def\bbbar{B\bar{B}}
\def\qqbar{q\bar{q}}
\def\chic{\chi_{c0}}
\def\kpkm{K^+K^-}

\def\pipi{\pi^+\pi^-}
\def\BF{{\cal{B}}}

\begin{document}
\vspace*{4cm}
\title{DALITZ ANALYSIS OF THREE-BODY B DECAYS}

\author{ A. GARMASH \\
(for the Belle Collaboration) }

\address{Department of Physics, Princeton University,\\
Princeton, NJ 08544, U.S.A.}

\maketitle\abstracts{
Results on Dalitz analysis of three-body charmless $\bpkpp$ and $\bpkkk$
decays are reported. We also present preliminary results on the studies of
direct CP violation in three-body decay $\bckpp$. The analysis is performed
using a large data sample collected with the Belle detector.}

\section{Introduction}

Studies of $B$ meson decays to three-body charmless hadronic final states are
a natural extension of studies of decays to two-body charmless final states.
Multiple resonances occurring nearby in phase space will interfere and a full
amplitude analysis is required to extract correct branching fractions for the
intermediate quasi-two-body states. In addition to being a rich laboratory for
studying $B$ meson decay dynamics, three-body charmless hadronic final
states may provide new possibilities for $CP$ violation searches~\cite{hhh-cp}.

\section{Apparatus, Data Sample \& Event Selection}

This analysis is based on a 140\,fb$^{-1}$ data
sample (152 million $B\bar{B}$ pairs) used for Dalitz analysis and on a 
253\,fb$^{-1}$ data sample (274 million $B\bar{B}$ pairs) used for searches
for direct CP violation in decay $\bckpp$. The data are collected with the
Belle detector operating at the KEKB asymmetric-energy $e^+e^-$ collider with
a center-of-mass energy at the $\Upsilon(4S)$ resonance.

  The Belle detector~\cite{Belle} is a large-solid-angle magnetic spectrometer
based on a 1.5~T superconducting solenoid magnet. Charged particle tracking is
provided by a three-layer silicon vertex detector and a 50-layer central
drift chamber (CDC) that surround the interaction point.
Charged hadron identification is provided by $dE/dx$ measurements in the CDC,
an array of 1188 aerogel \v{C}erenkov counters (ACC), and a barrel-like array
of 128 time-of-flight scintillation counters (TOF); information from the three
subdetectors is combined to form a single likelihood ratio, which is then used
for pion, kaon and proton discrimination.
For charged kaon identification, we impose a requirement on the particle
identification variable that has 86\% efficiency and a 7\% fake rate from
misidentified pions. Charged tracks that are positively identified as electrons
or protons are excluded from the analysis.

The dominant background to studied processes is due to
$e^+e^-\to~\qqbar$ ($q = u, d, s$ and $c$ quarks) continuum events. This type
of background is suppressed using variables that characterize the event
topology. This allows to reject about 98\% (92\%) of the continuum background
in the $\bckpp$ ($\bckkk$) decay while retaining 36\% (70\%) of the signal.
A detailed description of the continuum suppression technique can be found
in Ref.~\cite{khh-dalitz} and references therein.
The background that originates from other $B$ meson decays is studied using a
large sample of Monte Carlo generated events. We find that the dominant
$\bbbar$ background in the $\kppp$ final state is due to
$B^+\to\bar{D}^0\pi^+$,
$\bar{D}^0\to K^+\pi^-$ and also $B^+\to J/\psi(\psi(2S))K^+$,
$J/\psi(\psi(2S))\to \mu^+\mu^-$ decays. We veto these events by requiring
$|M(K\pi)-M_D|>0.10$~GeV/$c^2$, 
$|M(\pi^+\pi^-)_{\mu^+\mu^-}-M_{J/\psi}|>0.07$~GeV/$c^2$ and
$|M(\pi^+\pi^-)_{\mu^+\mu^-}-M_{\psi(2S)}|>0.05$~GeV/$c^2$, where subscript
$\mu^+\mu^-$ indicates that the muon mass assignment was used for charged
tracks to calculate the two-particle invariant mass.
We also remove $B^+\to \bar{D}^0K^+$,
$\bar{D}^0\to\pi^+\pi^-$ signal by requiring
$|M(\pi^+\pi^-)-M_D|>15$~MeV/$c^2$. To suppress the
background due to $\pi/K$ misidentification, we also exclude candidates if
the invariant mass of any pair of oppositely charged tracks from the $B$
candidate is consistent with the $\bar{D}^0\to K^+\pi^-$ hypothesis within
15~MeV/$c^2$, regardless of the particle identification information.
The most significant background from charmless
$B$ decays is found to originate from $B^+\to \eta'K^+$ followed by
$\eta'\to\pi^+\pi^-\gamma$, $B^+\to\rho^0\pi^+$ and $B^0\to K^+\pi^-$
decays. These backgrounds cannot be removed and are taken into account when
fitting the data.

\begin{figure}[!t]
  \includegraphics[width=0.32\textwidth]{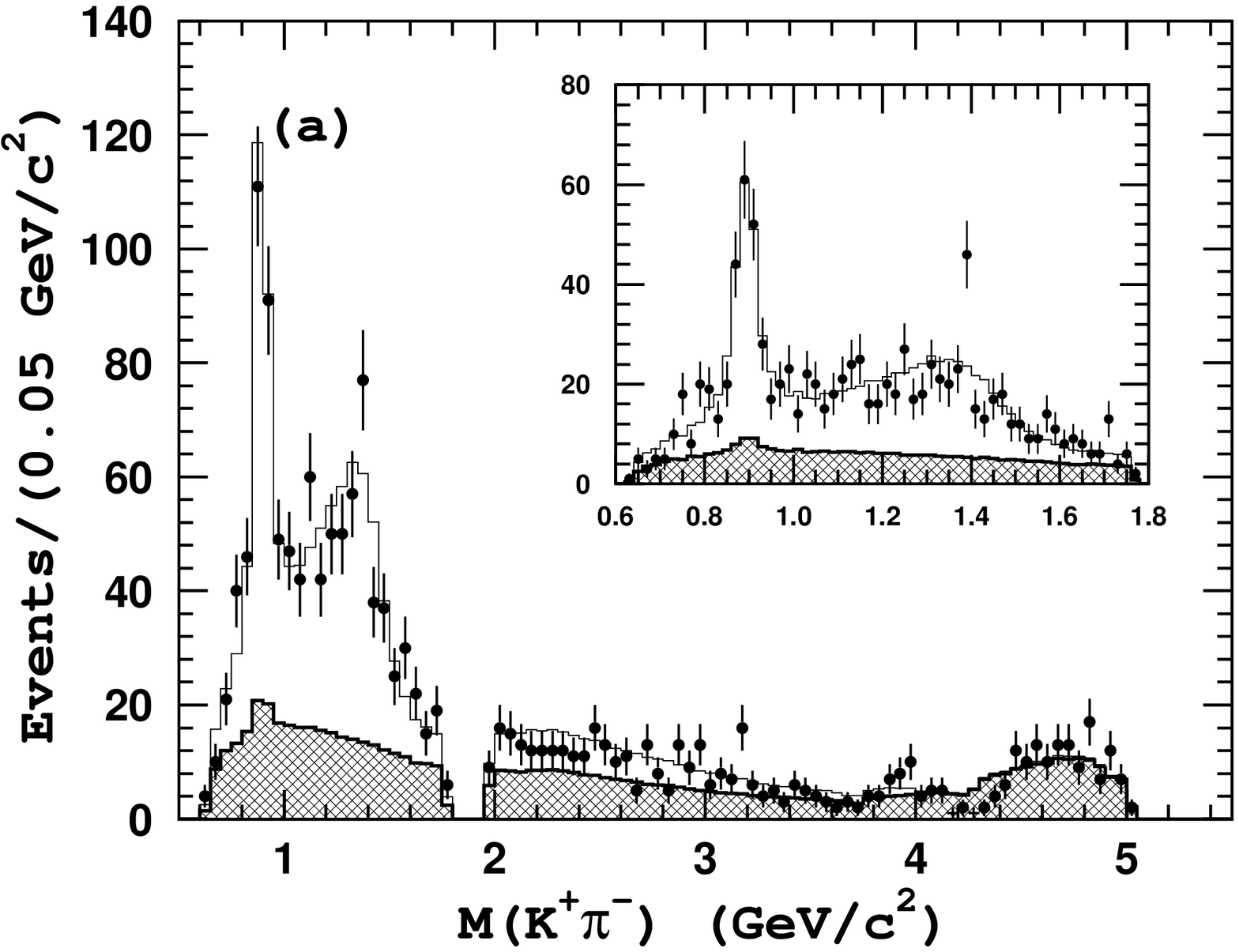} \hfill
  \includegraphics[width=0.32\textwidth]{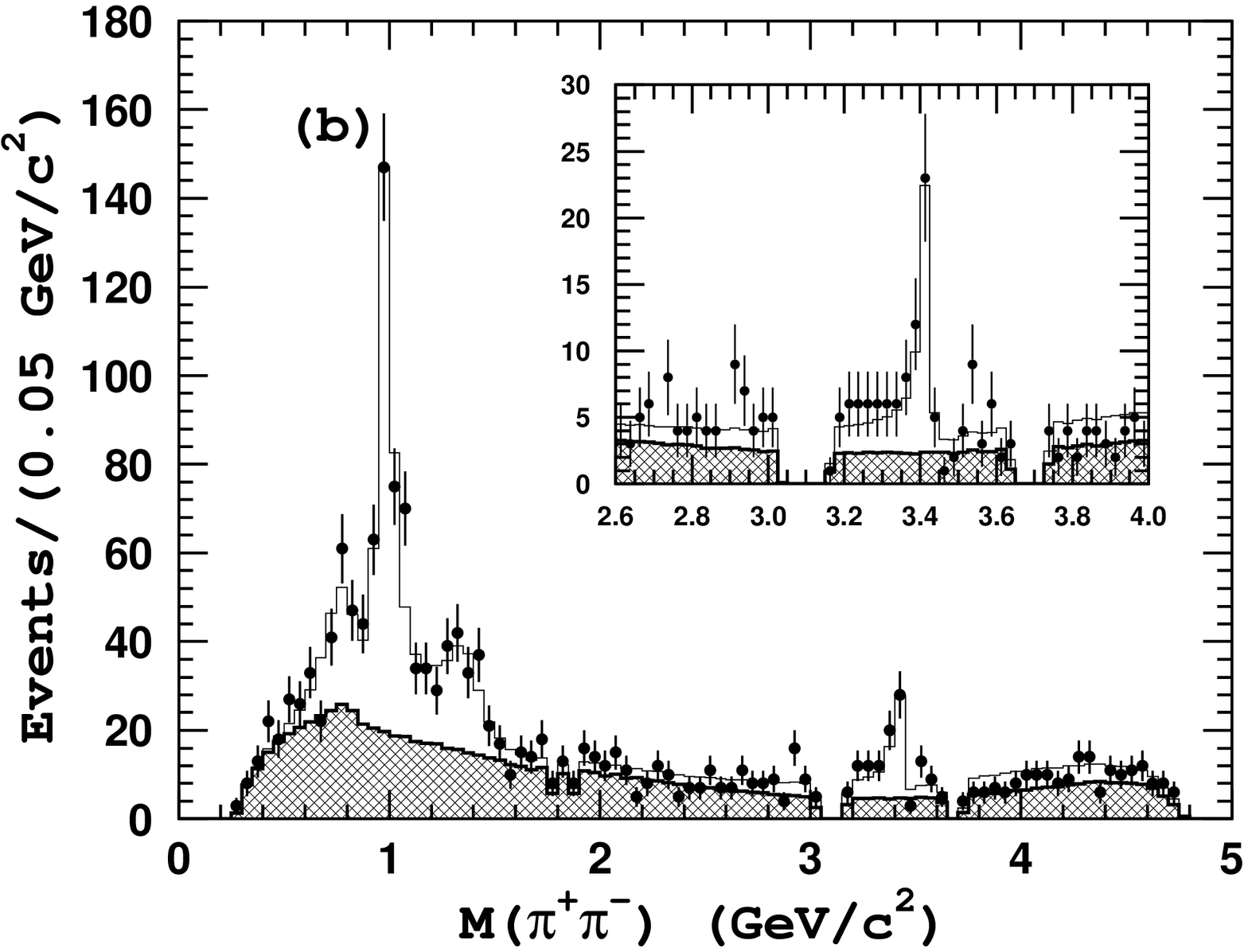} \hfill
  \includegraphics[width=0.32\textwidth]{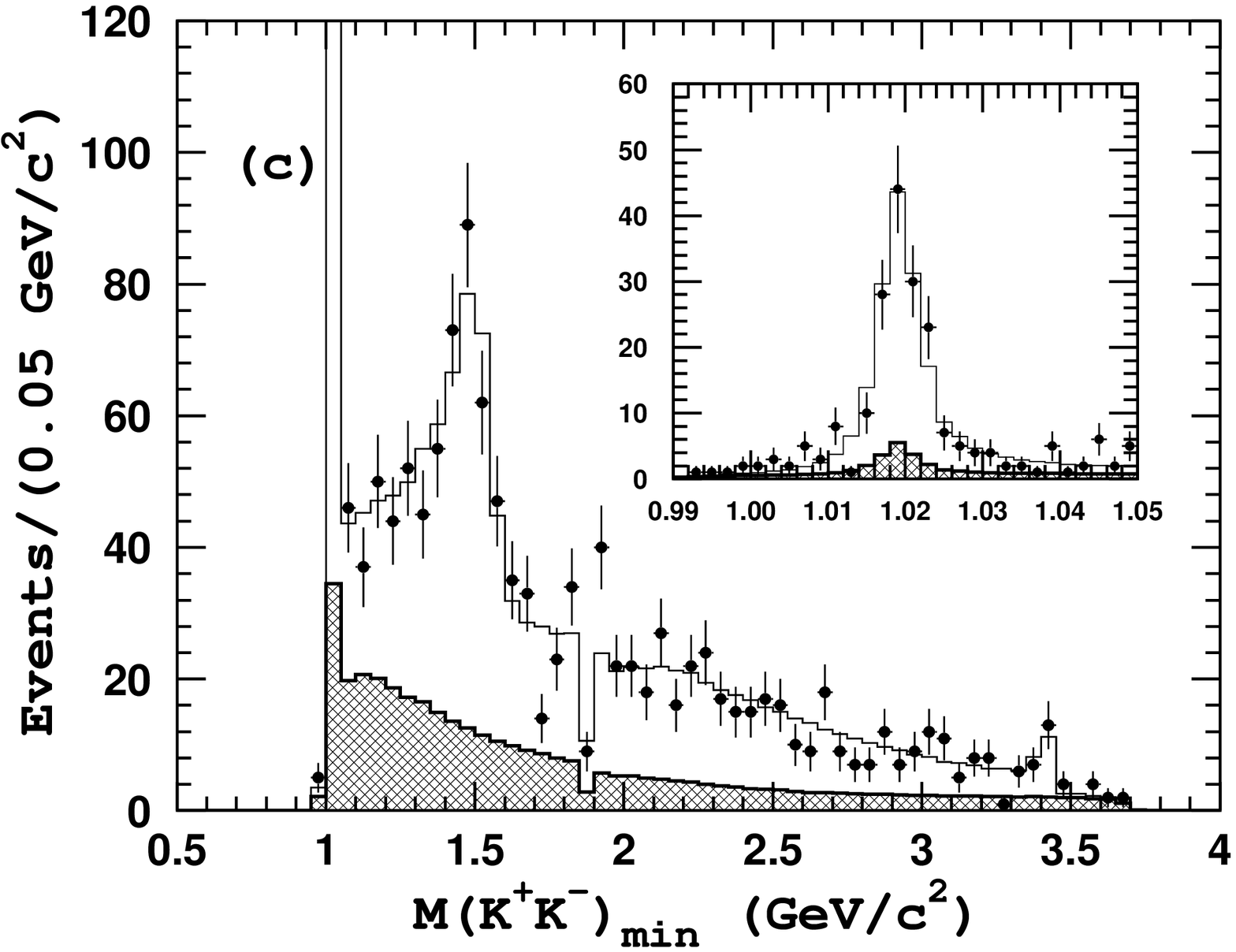}
\caption{Two-particle invariant mass spectra for events 
         in the $B$ signal region (open histograms) and for
         background events in the $\de-\mb$ sidebands (hatched histograms). 
         (a) $M(K^+\pi^-)$ spectrum with $M(\pi^+\pi^-)>1.5$~GeV/$c^2$;
         (b) $M(\pi^+\pi^-)$ with $M(K^+\pi^-)>2.0$~GeV/$c^2$~ and
         (c) $M(K^+K^-)_{min}$.}
\label{fig:hh-mass}
\end{figure}

The dominant background to the $\kpkk$ final state from other $B$ decays is
due to $B\to Dh$ process, where $h$ stands for a charged pion or kaon. To
suppress this background, we reject events where any two-particle invariant
mass is consistent with $\bar{D}^0\to K^+K^-$ or $\bar{D}^0\to K^+\pi^-$
within 15~MeV/$c^2$ regardless of the particle identification information.
We find no charmless $B$ decay modes that produce a significant background
to the $\kpkk$ final state.

\section{Results on Dalitz Analysis of $\bpkpp$ and $\bpkkk$}

Results of this analysis are described in detail in Ref.~\cite{khh-dalitz}.
There are 2584 $\kcpp$ and 1400 $\kckk$ events in the $B$ signal region that
satisfy all the selection requirements. Two-body invariant mass distributions
for these events are shown in Fig.~\ref{fig:hh-mass}. We find that the best fit
to $\kcpp$ events is obtained with a matrix element that is a coherent sum of
the  $K^*(892)^0\pi^+$, $K^*_0(1430)^0\pi^+$, $\rho(770)^0K^+$, $f_0(980)K^+$,
$f_X(1300)K^+$, $\chic K^+$ quasi-two-body channels and a non-resonant
component. A channel $f_X(1300)K^+$ (with mass and width of $f_X(1300)$ to be
determined from the fit) is added to account for an excess of signal events
visible in $M(\pipi)$ spectrum near $1.3$~GeV/$c^2$.
Results of the best fit are shown in Figs.~\ref{fig:hh-mass} (a,b).
The mass and width of the $f_X(1300)$ state obtained from the fit are
consistent with those for the $f_0(1370)$, however more data are
required for more definite conclusion.
To test for the contribution of other possible quasi-two-body intermediate
states such as $K^*(1410)^0\pi^+$, $K^*(1680)^0\pi^+$, $K^*_2(1430)^0\pi^+$
or $f_2(1270)K^+$, we include an additional amplitude of each of these
channels in the matrix element one by one and repeat the fit to data. None of
these channels have a statistically significant signal. Branching fraction
and upper limit results are summarized in Table~\ref{tab:branch}.

\begin{table}[!t]
  \caption{Summary of branching fraction results. The first quoted error is
           statistical, the second is systematic and the third is the model
           error.}
  \medskip
  \label{tab:branch}
\centering
  \begin{tabular}{|lcr|} \hline
\hspace*{20mm}Mode\hspace*{30mm} &
\hspace*{0mm}$\BF(B^+\to Rh^+)$ &
\hspace*{7mm}$\BF(B^+\to Rh^+)\times10^{6}$  \\
& \hspace*{0mm}$\times\BF(R\to h^+h^-)\times10^{6}$ & \\
 \hline
 $\kcpp$ charmless total & $-$
                        & $46.6\pm2.1\pm4.3$  \\
 $K^*(892)^0\pi^+$, $K^*(892)^0\to K^+\pi^-$
                        & $6.55\pm0.60\pm0.60^{+0.38}_{-0.57}$
                        & $9.83\pm0.90\pm0.90^{+0.57}_{-0.86}$     \\
 $K^*_0(1430)^0\pi^+$, $K^*_0(1430)^0\to K^+\pi^-$
                        & $27.9\pm1.8\pm2.6^{+8.5}_{-5.4}$
                        & $45.0\pm2.9\pm6.2^{+13.7}_{-~8.7}$       \\
%
 $K^*(1410)^0\pi^+$, $K^*(1410)^0\to K^+\pi^-$
                        & $<2.0$ & $-$   \hspace*{15mm}            \\
 $K^*(1680)^0\pi^+$, $K^*(1680)^0\to K^+\pi^-$
                        & $<3.1$ & $-$   \hspace*{15mm}            \\
 $K^*_2(1430)^0\pi^+$, $K^*_2(1430)^0\to K^+\pi^-$
                        & $<2.3$ & $-$   \hspace*{15mm}            \\
 $\rho(770)^0K^+$, $\rho(770)^0\to\pi^+\pi^-$
                        & $4.78\pm0.75\pm0.44^{+0.91}_{-0.87}$
                        & $4.78\pm0.75\pm0.44^{+0.91}_{-0.87}$     \\
 $f_0(980)K^+$, $f_0(980)\to\pi^+\pi^-$
                        & $7.55\pm1.24\pm0.69^{+1.48}_{-0.96}$
                        & $-$  \hspace*{15mm}                      \\
 $f_2(1270)K^+$, $f_2(1270)\to\pi^+\pi^-$
                        & $<1.3$ & $-$   \hspace*{15mm}            \\
 Non-resonant
                        & $-$
                        & $17.3\pm1.7\pm1.6^{+17.1}_{-7.8}$        \\
\hline
 $\kckk$ charmless total & $-$
                        & $30.6\pm1.2\pm2.3$                       \\
 $\phi K^+$, $\phi\to K^+K^-$
                        & $4.72\pm0.45\pm0.35^{+0.39}_{-0.22}$
                        & $9.60\pm0.92\pm0.71^{+0.78}_{-0.46}$     \\
 $\phi(1680)K^+$, $\phi(1680)\to K^+K^-$
                        & $<0.8$
                        & $-$      \hspace*{15mm}                  \\
 $f_0(980)K^+$, $f_0(980)\to K^+K^-$                
                        & $<2.9$
                        & $-$      \hspace*{15mm}                  \\
 $f'_2(1525)K^+$, $f'_2(1525)\to K^+K^-$
                        & $<4.9$
                        & $-$      \hspace*{15mm}                  \\
 $a_2(1320)K^+$, $a_2(1320)\to K^+K^-$ 
                        & $<1.1$
                        & $-$      \hspace*{15mm}                  \\
 Non-resonant
                        & $-$
                        & $24.0\pm1.5\pm1.8^{+1.9}_{-5.7}$         \\
\hline
 $\chic K^+$, $\chic\to\pi^+\pi^-$
                        & $1.37\pm0.28\pm0.12^{+0.34}_{-0.35}$
                        & $-$      \hspace*{15mm}                  \\
 $\chic K^+$, $\chic\to K^+K^-$
                        & $0.86\pm0.26\pm0.06^{+0.20}_{-0.05}$
                        & $-$      \hspace*{15mm}                  \\
 $\chic K^+$ combined   & $-$
                        & $196\pm35\pm33^{+197}_{-26}$             \\

\hline
  \end{tabular}
\end{table}

The best fit to $\kckk$ events is obtained with a matrix element that is a
coherent sum of the  $\phi K^+$, $f_X(1500)K^+$, $\chic K^+$ quasi-two-body
channels and a non-resonant component, where the $f_X(1500)K^+$ (with mass
and width of $f_X(1300)$ to be determined from the fit) channel is added to
describe the excess of signal events visible in $\kpkm$ mass spectrum near
$1.5$~GeV/$c^2$. As there are two identical kaons in the final state, the
decay amplitude is symmetrized with respect to interchange of two kaons of
the same charge. Results of the fit are shown in Fig.~\ref{fig:hh-mass}~(c)
and summarized in Table~\ref{tab:branch}. In order to check the sensitivity
of the data to the spin of the $f_X(1500)$ state, we replace the scalar
amplitude by a vector or a tensor amplitude for the $f_X(1500)$: the scalar
hypothesis gives the best fit. 
To estimate the possible contribution from other quasi-two-body channels, we
include an additional decay channel and repeat the fit to the data. In
particular we test the $\phi(1680)K^+$, $f'_2(1525)K^+$ and $a_2(1320)K^+$
channels. In all cases the fit finds no statistically
significant signal for the newly added channel. Since we observe a
clear $f_0(980)K^+$ signal in the $\kppp$ final state, we try to include
the $f_0(980)K^+$ amplitude in the $\bpkkk$ matrix element as well:
no statistically significant contribution from this channel is found.

\section{Search for Direct CP Violation in $\bckpp$}

\begin{figure}[!t]
  \centering
  \includegraphics[width=0.48\textwidth,height=52mm]{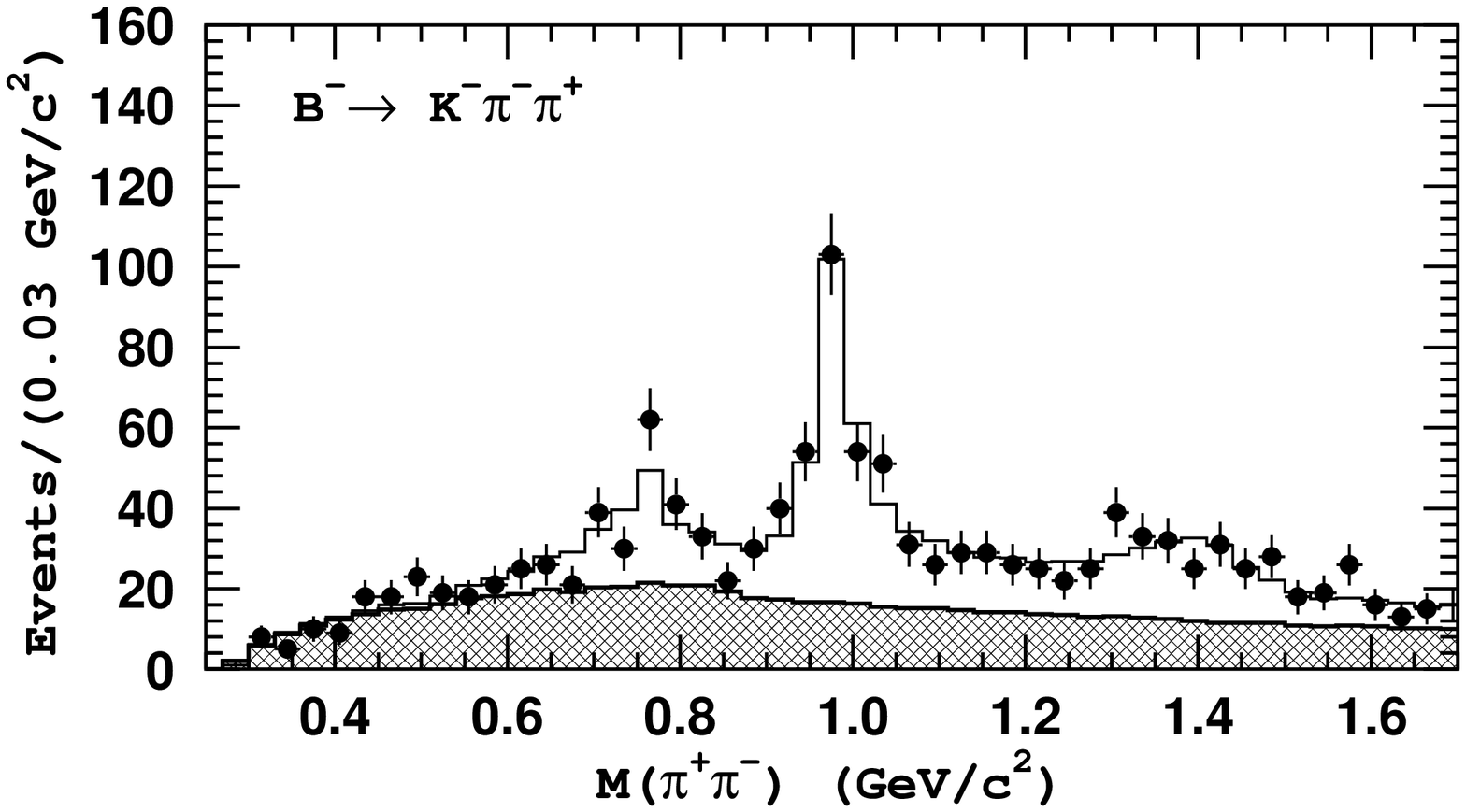} \hfill
  \includegraphics[width=0.48\textwidth,height=52mm]{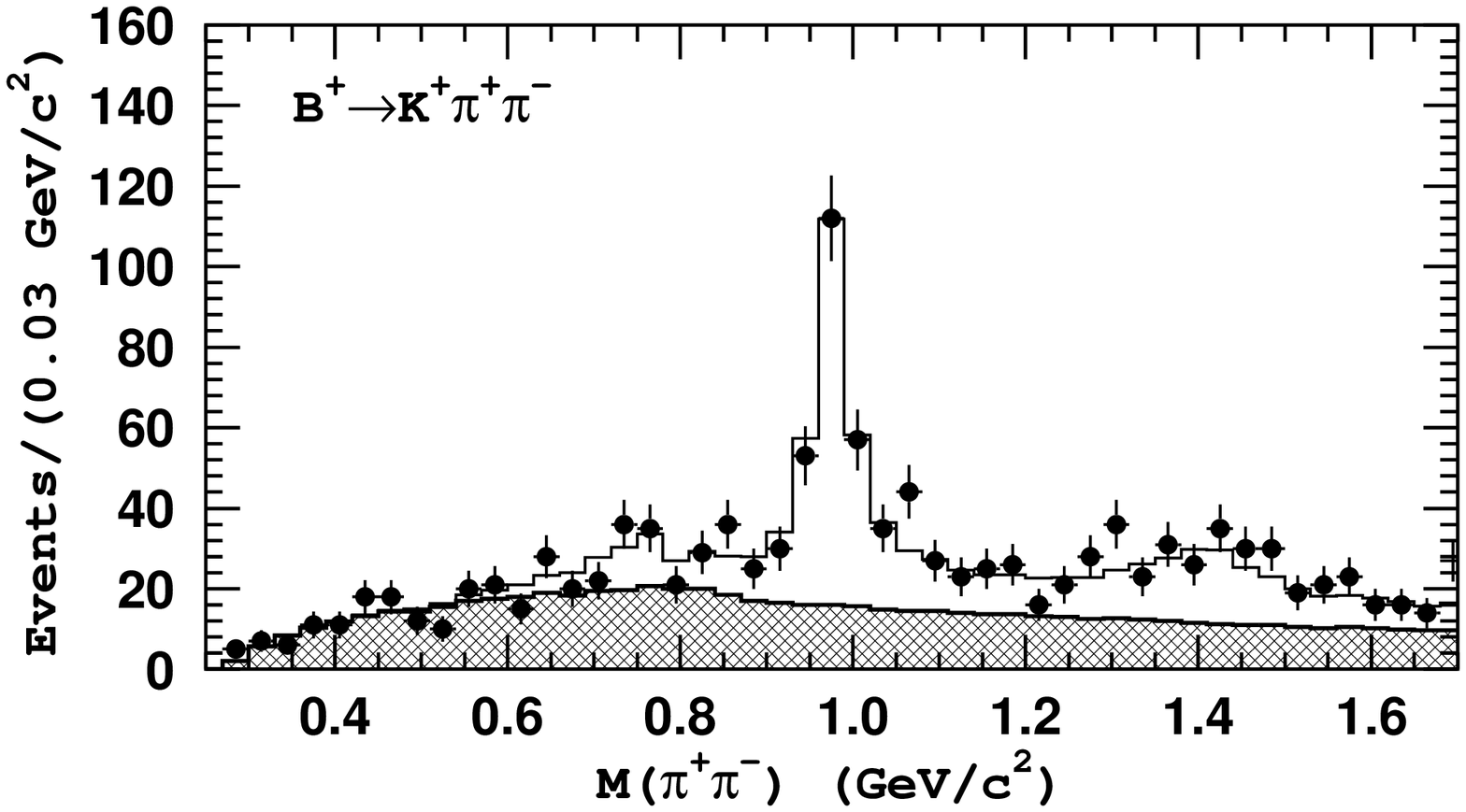} 
  \caption{$M(\pipi)$ mass spectra for $B^-$ (left) and $B^+$ (right).
           Points with error bars are data, the open histogram is the fit
           result and the hatched histogram is the
           background component.}
\label{fig:kpp-pp}
\end{figure}

In the analysis of the direct CP violation in $\bckpp$ decays we use
larger data sample of 253~fb$^{-1}$.
To search for direct CP violation we make a fit to $\kmpp$ and $\kppp$
samples separately. In the fit to $B^-$ and $B^+$ subsamples we fix
all the parameters not sensitive to CP violation (such as masses and
widths of resonance states) at values determined from the fit to the
combined $B^\pm$ sample. Results of calculations of CP violation asymmetries
defined as
\begin{equation}
\cal{A_{\rm CP}}=
          \frac{\BF{\it (B^-\to f^-)}-\BF{\it (B^+\to f^+)}}
               {\BF{\it (B^-\to f^-)}+\BF{\it (B^+\to f^+)}}
\end{equation}
are given in Table~\ref{tab:cp-results}.
In all except for $B^\pm\to\rho(770)^0K^\pm$ channels the measured asymmetry
agrees with zero within $1\sigma-1.5\sigma$ (after the model uncertainty is
accounted). A $2.4\sigma$ hint for a large direct CP violation is
observed in the decay $B^\pm\to\rho(770)^0K^\pm$. This is demonstrated in
Fig.~\ref{fig:kpp-pp}, where the $\pipi$ invariant mass distribution for the
$\rho(770)-f_0(980)$ mass region is shown. Analysis with a large data sample
is required in order to check this result.

\begin{table}[!h]
  \caption{Results of ${\cal{A}}_{\rm CP}$ calculation. Quoted errors are
           statistical, systematic and the model related error.}
  \medskip
  \label{tab:cp-results}
\centering
  \begin{tabular}{|lr|} \hline
\hspace*{20mm}Mode\hspace*{20mm} &
\hspace*{30mm}$\cal{A_{\rm CP}}$\hspace*{10mm} \\ \hline
 $\kcpp$ Charmless 
                        & $0.046\pm0.030\pm0.02$  \\
 $K^*(892)^0\pi^+$,  $~K^*(892)^0\to K^+\pi^-$ 
                        & $-0.14\pm0.08\pm0.02^{+0.03}_{-0.07}$     \\
 $K^*_0(1430)\pi^+$, $~K^*_0(1430)^0\to K^+\pi^-$
                        & $+0.06\pm0.05\pm0.02^{+0.01}_{-0.32}$     \\
 $\rho(770)^0K^+$,  $~\rho(770)^0\to\pi^+\pi^-$
                        & $+0.27\pm0.12\pm0.02^{+0.59}_{-0.03}$     \\
 $f_0(980)K^+$,  $~f_0(980)\to\pi^+\pi^-$
                        & $-0.13\pm0.11\pm0.02^{+0.16}_{-0.06}$     \\
 Non-resonant
                        & $-0.04\pm0.08\pm0.02^{+0.16}_{-0.03}$     \\
\hline
 $\chic K^+$, $\chic\to\pi^+\pi^-$
                        & $-0.30\pm0.22\pm0.02^{+0.42}_{-0.00}$     \\
\hline
  \end{tabular}
\end{table}

\end{document}